\documentclass[paper,notoc]{JHEP3}

\usepackage{epsfig}
\usepackage{graphicx}
\usepackage{feynmp}
\usepackage{amsmath}

\newcommand{\beq}{\begin{equation}}
\newcommand{\eeq}{\end{equation}}
\newcommand{\bea}{\begin{eqnarray}}
\newcommand{\eea}{\end{eqnarray}}
\newcommand{\nn}{\nonumber}

\def\eqn#1{Eq.~(\ref{#1})}
\def\eqns#1#2{Eqs.~(\ref{#1}) and~(\ref{#2})}

\def\fig#1{Fig.~{\ref{#1}}}
\def\sec#1{Section~{\ref{#1}}}

\def\bsp#1\esp{\begin{split}#1\end{split}}

\newcommand\fverb{\setbox\pippobox=\hbox\bgroup\verb}
\newcommand\fverbdo{\egroup\medskip\noindent%
                       \fbox{\unhbox\pippobox}\ }
\newcommand\fverbit{\egroup\item[\fbox{\unhbox\pippobox}]}
\newbox\pippobox

\def\cM{{\cal M}}

\newcommand{\epslc}{\epsilon_{1234}}
\newcommand{\eps}{\epsilon}
\newcommand{\li}{\textrm{Li}}
\newcommand{\re}{\textrm{Re}~}
\newcommand{\im}{\textrm{Im}~}
\newcommand{\br}[1]{\langle #1 \rangle}
\newcommand{\sq}[1]{[#1]}

\newcommand{\ord}{\begin{cal}O\end{cal}}

\newcommand{\muterm}[1]{\left(\frac{\mu^2}{#1}\right)^\eps}

\def\cg{c_\Gamma}
\newcommand\sss{\scriptscriptstyle}
\newcommand\as{\alpha_{\sss S}} 
\newcommand\gs{g}

\def\tgs{\bar\gs}
\def\bC{\bar C}
\def\bV{\bar V}

\def\balpha{\bar\alpha}
\def\vc{{\rm V\! C}}
\newcommand{\ph}[2]{(#1)_{#2}}

\title{The five-gluon amplitude in the high-energy limit}

\author{Vittorio Del Duca\\
Istituto Nazionale di Fisica Nucleare\\
Laboratori Nazionali di Frascati\\
00044 Frascati (Roma), Italy\\
       E-mail: \email{delduca@lnf.infn.it}}

\author{Claude Duhr\\
Institut de Physique Th\'eorique \&
Centre for Particle Physics and Phenomenology (CP3)\\
Universit\'e Catholique de Louvain\\
Chemin du Cyclotron 2,
B-1348 Louvain-la-Neuve, Belgium\\
E-mail: \email{claude.duhr@uclouvain.be}}

\author{E.~W.~N.~Glover\\
Institute for Particle Physics Phenomenology, 
University of Durham\\ Durham, DH1 3LE, U.K.\\
E-mail: \email{E.W.N.Glover@durham.ac.uk}}

\abstract{We consider the high energy limit of the colour ordered one-loop
five-gluon amplitude in the planar maximally supersymmetric
$\begin{cal}N\end{cal}=4$ Yang-Mills theory in the multi-Regge kinematics where
all of the gluons are strongly ordered in rapidity. We apply the calculation of
the one-loop pentagon in $D=6-2\eps$ performed in a companion paper~\cite{noi}
to compute the one-loop five-gluon amplitude through to ${\cal O}(\eps^2)$. 
Using the factorisation properties of the amplitude in the high-energy limit, we
extract the one-loop gluon-production vertex to the same accuracy,  and, by
exploiting the iterative structure of the gluon-production vertex implied by the
BDS ansatz, we perform the first computation of the two-loop gluon-production
vertex up to and including finite terms.}

\preprint{CP3-09-17\\ IPPP/09/34}

\begin{document}

\section{Introduction}

The colour-stripped one-loop five-gluon MHV amplitude
in the planar maximally supersymmetric $\begin{cal}N\end{cal}=4$ Yang-Mills theory
to all orders in $\eps$ is given by~\cite{Bern:1996ja,Bern:2006vw},
\beq
m_5^{(1)}(1,2,3,4,5) =  - \frac{1}{4} \sum_{\rm cyclic} s_{12} s_{23} I_4^{1m}(1,2,3,45,\eps) - 
\frac{\eps}{2} \eps_{1234} I_5^{6-2\eps}(\eps)\, ,\label{eq:pent}
\eeq
%
where $m_5^{(1)}$ denotes the one-loop coefficient, \emph{i.e.}, the one-loop amplitude rescaled by the tree-level amplitude, and where the cyclicity is over $i=1,\ldots,5$. Here $I_4^{1m}(1,2,3,45,\eps)$ represents the one-mass box integral
with an off-shell leg of virtuality $s_{45}$, $I_5^{6-2\eps}(\eps)$
is the (massless) one-loop pentagon integral evaluated in $6-2\eps$ dimensions, and the contracted Levi-Civita tensor is
$\eps_{1234}= {\rm tr}[\gamma_5\!\!\not\!k_1\!\!\not\!\!k_2\!\not\!\!k_3\!\!\not\!k_4]$.
In a companion paper~\cite{noi}, we have performed the first analytic computation of
the higher dimension pentagon, $I_5^{6-2\eps}(\eps)$, albeit in the simplified kinematical set-up of the 
{\em multi-Regge kinematics}~\cite{Kuraev:1976ge}. In this limit, we have derived $I_5^{6-2\eps}(\eps)$
as an all-order expression in $\eps$, and explicitly expanded it through to $\ord(\eps^2)$. 

In the high-energy limit (HEL) $s\gg |t|$, any scattering process is dominated
by the exchange of the highest-spin particle in the crossed channel. 
Thus, in perturbative QCD the leading contribution in powers of $s/t$
to any scattering process comes from gluon exchange in the $t$ channel.
In this limit, scattering amplitudes undergo a Regge factorisation~\cite{Kuraev:1976ge,Fadin:1993wh} 
which allows an amplitude for gluon exchange to be decomposed in terms of building blocks associated with the various
components of the amplitude. In the simplest case of four-gluon scattering, one exchanges a reggeised gluon (representing a gluon ladder) that is emitted from one scattering vertex and absorbed at the other.   This emission is described by the coefficient function or impact factor.   For processes involving more gluons, additional gluons are emitted by the gluon ladder and this emission is controlled by the gluon-production (or Lipatov) vertex. Using the high-energy factorisation for 
colour-stripped amplitudes~\cite{DelDuca:2008pj,DelDuca:2008jg}, we can relate
the one-loop gluon-production vertex in the
$\begin{cal}N\end{cal}=4$ super Yang-Mills theory to the one-loop five-gluon amplitude given in~\eqn{eq:pent} and extract it through to $\ord(\eps^2)$.

Recently, Bern, Dixon and Smirnov (BDS) have proposed an iterative
ansatz~\cite{Anastasiou:2003kj,Bern:2005iz}   for the $l$-loop  $n$-gluon
scattering amplitude in the maximally supersymmetric $\begin{cal}N\end{cal}=4$  Yang-Mills theory
(MSYM), with the maximally-helicity violating (MHV) configuration and for 
arbitrary $l$ and $n$. The iterative structure of the BDS ansatz  has been shown
to be correct for the two-loop five-point amplitude through direct numerical
calculation~\cite{Bern:2006vw}. Together with the high-energy factorisation,
that implies an iterative structure of the gluon-production
vertex~\cite{DelDuca:2008jg}. Thus, the knowledge of the one-loop
gluon-production vertex through to $\ord(\eps^2)$, allows us to perform the
first computation of the two-loop gluon-production vertex up to and including
the finite terms.

Our paper is organised as follows.  In \sec{sec:mrkforall} we consider the
five-point amplitude in the multi-Regge kinematics.   First we make a precise
definition of the multi-Regge kinematics and review how the tree-level MHV
amplitude factorises in the high energy limit. The factorisation properties of
the five-gluon amplitude are described in \sec{sec:5pthel} and the relationship
between the one-loop amplitude and the one-loop Lipatov vertex established.   We
remind the reader of the iterative structure of the Lipatov vertex in
\sec{sec:iterative} while its analytic continuation properties are discussed in
\sec{sec:analytic}.  In \sec{sec:1loop5ptamp} we present the one-loop five-point
amplitude through to $\ord(\eps^2)$ and use it to compute the one-loop
gluon-production vertex through to $\ord(\eps^2)$ in \sec{sec:onelooplip} where we find contributions from both the parity-even and parity-odd parts.  In
\sec{sec:twolooplip} we compute the two-loop gluon-production vertex through to
finite terms.  Our findings are briefly summarised in \sec{sec:concl}.  Some of
the technical details are enclosed in the Appendices. Further details on the
multi-parton light-cone momenta and how they behave in the multi-Regge
kinematics are given in Appendices A and B, while the soft limit of the one-loop
Lipatov vertex is further discussed in Appendix C.

\section{Five-point amplitudes in multi-Regge kinematics}
\label{sec:mrkforall}

We consider a five-point amplitude, $g_1\,g_2\to g_3\,g_4\, g_5$, with
all the momenta taken as outgoing, and label the legs cyclically clockwise. In the 
multi-Regge kinematics~\cite{Kuraev:1976ge}, the produced particles are
strongly ordered in rapidity and have comparable transverse momenta,
\begin{equation}
y_3 \gg y_4\gg y_5;\qquad |p_{3\perp}| \simeq |p_{4\perp}| \simeq|p_{5\perp}|\, 
.\label{mrknpt}
\end{equation}
Accordingly, the Mandelstam invariants can be written in the approximate 
form (\ref{mrinv})\footnote{A physically more intuitive representation of the invariants in terms of rapidities
is given in Ref.~\cite{DelDuca:2008jg}.}.
%
%
We label the momenta transferred in the $t$-channel as
\beq
q_1 = p_1+p_5\, ,\qquad
q_2 = -p_2-p_3\, ,\label{eq:mome}
\eeq
with virtualities $t_i = q_i^2$.
Then it is easy to see that in the multi-Regge kinematics the transverse
components of the momenta $q_i$ dominate over the longitudinal components,
$q_i^2 \simeq - |q_{i\perp}|^2$. In addition, $t_1=s_{15}$ and
$t_2=s_{23}$, and we label $s = s_{12}$, and $s_1=s_{45}$,
$s_2=s_{34}$.
Thus, the multi-Regge kinematics (\ref{mrknpt}) become
\begin{equation}
s \gg s_{1},\ s_{2} \gg -t_1,\ -t_2\, .\label{eq:mrknpt2} 
\end{equation}
Labelling the transverse momentum of the particle emitted along the ladder as
$\kappa=|p_{4\perp}|^2$, we can write
\begin{equation}
\kappa = \frac{s_{1}\, s_{2}}{s}
,\label{massnpt}
\end{equation}
which is known as the mass-shell condition (\ref{eq:masshell}).
%

\subsection{Tree amplitudes in multi-Regge kinematics}
\label{sec:mhvmrk}

\begin{figure}[!t]
\begin{center}
 \begin{fmffile}{mr2}
                  \begin{fmfgraph*}(100,80)
                \fmfstraight
                   \fmfleft{p1,pd4,p2}
                   \fmfright{x1,x6,x7}
                   \fmf{phantom}{p1,u1,v1n,u2,pn,u3,x1}
                   \fmf{phantom}{p2,o1,v23,o2,p3,o3,x7}
                   \fmffreeze
                   \fmf{phantom}{pn,p4,p3}
                   \fmffreeze
                   \fmf{gluon,label=$p_2$,label.side=left,l.d=0.03w}{p2,v23}
                   \fmf{gluon,label=$p_3$,label.side=left,l.d=0.03w}{v23,p3}
                   \fmf{phantom}{v23,v4,v1n}
                   \fmffreeze
                   \fmfv{decor.shape=circle,decor.filled=shaded,decor.size=.09w,fore=green}{v23}
                   \fmfv{decor.shape=circle,decor.filled=shaded,decor.size=.09w,fore=green}{v4}
                   \fmfv{decor.shape=circle,decor.filled=shaded,decor.size=.09w,fore=green}{v1n}
                   \fmf{zigzag,label=$q_2$,label.side=right,l.d=0.055w}{v23,v4}
                   \fmf{zigzag,label=$q_{1}$,label.side=right,l.d=0.055w}{v4,v1n}
                   \fmf{gluon,label=$p_1$,label.side=left,l.d=0.03w}{p1,v1n}
                   \fmf{gluon,label=$p_5$,label.side=left,l.d=0.03w}{v1n,pn}
                   \fmffreeze
                   \fmf{gluon,label=$p_4$,label.side=left,l.d=0.03w}{v4,p4}
                   \fmffreeze
                   \fmf{phantom}{u3,oun1,o3}
                   \fmffreeze
                   \fmf{phantom}{o3,ox1,ox2,ox3,ox4,ox5,ox6,ox7,ox8,ox9,ox10,oun1}
                   \fmffreeze
                   \fmf{plain,tension=0.2,left=0.3,label=$s_2$}{ox1,ox10}
                   \fmffreeze
                   \fmf{phantom}{oun1,ou4x1,ou4x2,ou4x3,ou4x4,ou4x5,ou4x6,ou4x7,ou4x8,ou4x9,ou4x10,u3}
                   \fmffreeze
                   \fmf{plain,tension=0.2,left=0.3,label=$s_1$}{ou4x1,ou4x10}
                   \fmffreeze
                    \fmfv{label=$\kappa$,l.a=-180,l.d=-0.14w}{p4}
                         \end{fmfgraph*}
                            \end{fmffile}
                            \end{center}
\caption{\label{fig:MR} Five-point amplitude in the multi-Regge kinematics. 
The green blobs indicate the coefficient functions (impact factors) and the vertex describing the 
emission of gluons along the ladder.}
\end{figure}
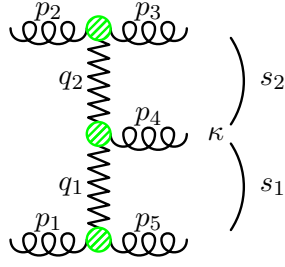

The colour decomposition of the tree-level five-point amplitude
is~\cite{Mangano:1990by}
\begin{equation}
\cM_5^{(0)}(1,2,3,4,5) = 2^{5/2}\, \sum_{S_5/Z_5} {\rm tr}(T^{d_1}
\cdots T^{d_5}) \, m_5^{(0)}(1,2,3,4,5)\, ,\label{one}
\end{equation}
where $d_i$ is the colour of a gluon of momentum $p_i$ and helicity $\nu_i$.
The $T$'s are the colour matrices\footnote{We use the normalization
${\rm tr}(T^c T^d) = \delta^{cd}/2$, although it is immaterial in what follows.} in the
fundamental representation of SU($N$) and the sum is over the noncyclic
permutations $S_5/Z_5$ of the set $[1, \ldots ,5]$. For five gluons, there are only MHV
helicity configurations $(-,-,+,+,+)$ for which the tree-level gauge-invariant 
colour-stripped amplitudes assume the form
\begin{equation}
m_5^{(0)}(1,2,3,4,5) = g^3\, \frac{\langle p_i p_j\rangle^4}
{\langle p_1 p_2\rangle \langle p_2 p_3\rangle \langle p_3 p_4\rangle
\langle p_4 p_5\rangle \langle p_5 p_1\rangle}\, ,\label{two}
\end{equation}
where $i$ and $j$ are the two gluons of negative helicity. The colour structure
of \eqn{one} in multi-Regge kinematics is 
known~\cite{DelDuca:1993pp,DelDuca:1995zy,DelDuca:1999rs}
and will not be considered further.
Here we concentrate on the behaviour of the colour-stripped 
amplitude (\ref{two}). Using the spinor products in multi-Regge kinematics~(\ref{mrpro}),
the amplitude (\ref{two}) takes the factorised form~\cite{DelDuca:1995zy},
\begin{equation}
m_5^{(0)}(1,2, \ldots ,5) = s \left[g\, C^{(0)}(p_2,p_3) \right]\, 
\frac{1}{t_2}\, \left[g\,V^{(0)}(q_2,q_1;\kappa)\right] \frac{1}{ t_1}\, 
\left[g\, C^{(0)}(p_1,p_5) \right]\, \label{treenpt}
\end{equation}

which is shown schematically in~\fig{fig:MR}.

The gluon coefficient functions $C^{(0)}$, which yield the LO gluon impact factors, 
are given in Ref.~\cite{Kuraev:1976ge} in terms of their spin structure 
and in Ref.~\cite{DelDuca:1995zy,DelDuca:1996km} at fixed
helicities of the external gluons,
\begin{equation}
C^{(0)}(p_2^-,p_3^+) = 1 \qquad C^{(0)}(p_1^-,p_5^+) = \frac{p_{5\perp}^*} 
{p_{5\perp}}\, ,\label{centrc}
\end{equation}
with the complex transverse momentum $p_{\perp}=p_x+ip_y$.
The vertex for the emission of a gluon along the ladder 
is~\cite{DelDuca:1995zy,Lipatov:1976zz,Lipatov:1991nf}
\begin{equation}
V^{(0)}(q_2,q_1,\kappa) = \sqrt{2}\, \frac{q^*_{2\perp} q_{1\perp}}{p_{4\perp}}\, 
,\label{lipeq}
\end{equation}
with $p_4 = q_2 - q_1$.

\subsection{High-energy factorisation of the five--point amplitude}
\label{sec:5pthel}

The Regge factorisation of the
tree-level colour-stripped amplitude is given by \eqn{treenpt}.
In the leading logarithmic (LL) approximation,
the virtual radiative corrections to Eq.~(\ref{treenpt}) are obtained, to all orders
in $\as$, by replacing the propagator of the $t$-channel gluon by its
reggeised form~\cite{Kuraev:1976ge}. That is, by making the replacement
\begin{equation}
\frac{1}{ t_i} \to \frac{1}{t_i} 
\left(\frac{s_i}{ \tau}\right)^{\alpha(t_i)}\, ,\label{sud}
\end{equation}
in Eq.~(\ref{treenpt}), where $\alpha(t_i)$ can be written in
dimensional regularization in $d=4-2\epsilon$ dimensions as
\begin{equation}
\alpha(t_i) = \gs^2\, c_{\Gamma}\,  
\left(\frac{\mu^2}{ -t_i}\right)^{\epsilon} \, N\, \frac{2}{\epsilon}
,\label{alph}
\end{equation}
with $N$ colours, and
\begin{equation}
c_{\Gamma} = \frac{1}{(4\pi)^{2-\epsilon}}\, \frac{\Gamma(1+\epsilon)\,
\Gamma^2(1-\epsilon)}{ \Gamma(1-2\epsilon)}\, .\label{cgam}
\end{equation}
$\alpha(t_i)$ is the Regge trajectory and accounts for the
higher order corrections to gluon exchange in the $t_i$ channel. In \eqn{sud},
the reggeisation scale $\tau$ is introduced to separate contributions to
the 
reggeized propagator, the coefficient
function and the gluon-production vertex.  It is much smaller than any of the $s$-type
invariants, and it is of the order of the $t$-type invariants.
In order to go beyond the LL approximation and to compute the higher-order
corrections to the gluon-production vertex (\ref{lipeq}), we need a high-energy
prescription~\cite{Fadin:1993wh} which
disentangles the virtual corrections to the gluon-production vertex
from those to the coefficient functions (\ref{centrc})
and from those that reggeize the gluon (\ref{sud}). 
The high-energy prescription of Ref.~\cite{Fadin:1993wh} is given 
at the colour-dressed amplitude level in QCD, where it holds
to the next-to-leading-logarithmic (NLL) accuracy. In Ref.~\cite{DelDuca:2008pj},
we showed that the high-energy prescription, when applied to the colour-stripped four-point
amplitude, is valid up to three loops. In Ref.~\cite{DelDuca:2008jg},
we conjectured the factorised form of a generic colour-stripped $n$-gluon amplitude
in the multi-Regge kinematics. 
For the five-point amplitude, $g_1\,g_2\to g_3\,g_4\,g_5$, that 
prescription yields
\beq
m_5 = s \left[g\, C(p_2,p_3,\tau) \right]\, 
{1\over t_2}\, \left({-s_2\over \tau}\right)^{\alpha(t_2)}
\left[g\,V(q_2,q_1,\kappa,\tau)\right]\, 
{1\over t_1}\, \left({-s_1\over \tau}\right)^{\alpha(t_1)}
\left[g\, C(p_1,p_5,\tau) \right]\, , \label{loop5pt}
\eeq
with the invariants labelled as in \sec{sec:mrkforall}, {\it i.e.}
$t_1=s_{51}$, $t_2=s_{23}$, $s_1=s_{45}$ and $s_2=s_{34}$.
In \eqn{loop5pt}, we suppressed the dependence of the coefficient function and of the 
gluon-production vertex on the dimensional regularisation parameters $\mu^2$ and $\eps$.
In order for the amplitude $m_5$ to be real,
\eqn{loop5pt} is taken in the Euclidean region where all the invariants are negative,
\beq
s, s_{1}, s_{2}, t_1, t_2, \kappa < 0\, .\label{eq:neginv}
\eeq
Thus, the multi-Regge kinematics (\ref{eq:mrknpt2}) become,
\begin{equation}
-s \gg -s_{1}, -s_{2} \gg -t_1, -t_2\, .\label{eq:mrk2} 
\end{equation}
Then the mass-shell condition (\ref{massnpt}) for the intermediate gluon is
\begin{equation}
- \kappa = {(-s_{1})\, (-s_{2})\over -s}\, ,\label{mass}
\end{equation}
where $\kappa= - |p_{4\perp}|^2$.

In \eqn{loop5pt}, the Regge trajectory has the perturbative expansion,
\begin{equation}
\alpha(t_i) = \tgs^{2} \balpha^{(1)}(t_i) + \tgs^4 \balpha^{(2)}(t_i) + 
\tgs^6 \balpha^{(3)}(t_i) + \ord(\tgs^8)\, ,\label{alphb}
\end{equation}
with $i=1,2$, and with the rescaled coupling
\beq
\tgs^2 = \gs^2 \cg N\, .\label{rescal}
\eeq
In \eqn{loop5pt}, the coefficient functions $C$ and the gluon-production vertex $V$ are also
expanded in the rescaled coupling,
\begin{eqnarray} 
C(p_i,p_j,\tau) &=& C^{(0)}(p_i,p_j)\left(1 + \sum_{r=1}^{s-1} \tgs^{2r} \bC^{(r)}(t_k,\tau) 
+ \ord(\tgs^{2s}) \right)\, 
,\label{fullv} \\
V(q_2,q_1,\kappa,\tau) &=& V^{(0)}(q_2,q_1)\left(1 + \sum_{r=1}^{s-1} \tgs^{2r} 
\bV^{(r)}(t_1,t_2,\kappa,\tau) + \ord(\tgs^{2s}) \right)\, .\nn
\end{eqnarray}
with $(p_i+p_j)^2=t_k$ where $C$ and $V$ are real,
up to overall complex phases in $C^{(0)}$, \eqn{centrc}, and $V^{(0)}$, 
\eqn{lipeq}, induced by the complex-valued helicity bases. Note that because several transverse scales 
occur, we prefer to associate the renormalisation scale dependence of the trajectory, coefficient function and gluon-production vertex with the loop coefficients rather than in the rescaled coupling,
\bea
&& \balpha^{(n)}(t_i) = \left({\mu^2\over -t_i}\right)^{n\eps} \alpha^{(n)}\, 
,\quad \bC^{(n)}(t_k,\tau) = \left({\mu^2\over -t_k}\right)^{n\eps} 
C^{(n)}(t_k,\tau)\, ,\nn\\
&& \bV^{(n)}(t_1,t_2,\kappa,\tau) = \left({\mu^2\over -\kappa}\right)^{n\eps} 
V^{(n)}(t_1,t_2,\kappa,\tau)\, .\label{eq:coeffrescal}
\eea

The perturbative expansion of \eqn{loop5pt} can be written as
\beq
m_5 = m_5^{(0)} \left( 1 + \tgs^2\ m_5^{(1)} + \tgs^4 m_5^{(2)} + \tgs^6 m_5^{(3)} 
+ \ord(\tgs^8) \right)\, .\label{elasexpand}
\eeq
In the expansion of \eqn{elasexpand}, the knowledge of
the $l$-loop five-point amplitude in the multi-Regge kinematics (\ref{eq:mrk2}),
together with the $l$-loop trajectory $\alpha^{(l)}$ and coefficient function $C^{(l)}$,
allows one to derive the gluon-production vertex to the same accuracy.
The one-loop coefficient is
\beq
m_5^{(1)}(\eps) = \balpha^{(1)}(t_1) L_1 + \balpha^{(1)}(t_2) L_2
+ \bC^{(1)}(t_1,\tau) + \bC^{(1)}(t_2,\tau) + \bV^{(1)}(t_1,t_2,\kappa,\tau)\, 
.\label{exp1loop}
\eeq
where $L_i=\ln(-s_i/\tau)$ and $i=1,2$. The one-loop trajectory is~\cite{Kuraev:1976ge},
\beq
\alpha^{(1)} = \frac{2}{\eps}\, ,\label{eq:alph1}
\eeq
and the one-loop coefficient function is~\cite{Fadin:1993wh,DelDuca:1998kx,DelDuca:1998cx,Bern:1998sc,Fadin:1992zt,Fadin:1993qb} to all orders in $\eps$ given by
\beq\label{eq:C1l}
C^{(1)}(t,\tau)=\frac{\psi(1+\eps)-2\psi(-\eps)+\psi(1)}{\eps}
-\frac{1}{\eps}\,\ln\frac{-t}{\tau}.
\eeq
Since $\alpha^{(1)}$ and $C^{(1)}(t,\tau)$ are known to all orders in $\eps$,  we see that the order to which $m_5^{(1)}(\eps)$ is known dictates the order to which one may extract $V^{(n)}(t_1,t_2,\kappa,\tau)å$.

Similarly the two-loop coefficient of the five-point amplitude is
\bea\label{eq:twoexp}
m_5^{(2)}(\eps) &=& \frac{1}{2} \left[ m_5^{(1)}(\eps) \right]^2
+ \balpha^{(2)}(t_1) L_1 + \balpha^{(2)}(t_2) L_2 \label{exp2loop}\\
&+& \bC^{(2)}(t_1,\tau)  + 
\bV^{(2)}(t_1,t_2,\kappa,\tau) + \bC^{(2)}(t_2,\tau)\nn\\
&-& \frac{1}{2} \left( \bC^{(1)}(t_1,\tau) \right)^2
- \frac{1}{2} \left( \bV^{(1)}(t_1,t_2,\kappa,\tau) \right)^2
- \frac{1}{2} \left( \bC^{(1)}(t_2,\tau) \right)^2
\, ,\nn
\eea
where $m_5^{(1)}(\eps)$, $\bC^{(1)}(t,\tau)$ and $ \bV^{(1)}(t_1,t_2,\kappa,\tau)$  must be known 
through to $\ord(\eps^2)$.   The two-loop trajectory, $\alpha^{(2)}$, is known in full
QCD~\cite{Fadin:1995xg,Fadin:1995km,Fadin:1996tb,Blumlein:1998ib,DelDuca:2001gu}.
In MSYM, it has been computed through $\ord(\eps^0)$ 
directly~\cite{Kotikov:2000pm} and using the maximal
trascendentality principle~\cite{Kotikov:2002ab},  and through 
$\ord(\eps^2)$ directly~\cite{DelDuca:2008pj},
\beq
\alpha^{(2)} = - {2\zeta_2\over\eps} - 2\zeta_3 - 8\zeta_4\eps
+ (36\zeta_2\zeta_3 + 82\zeta_5)\eps^2 + \ord(\eps^3)\, .
\label{eq:tworegge}
\eeq
The MSYM two-loop coefficient function has been computed
through $\ord(\eps^2)$~\cite{DelDuca:2008pj},
\bea
C^{(2)}(t,\tau) &=& \frac{2}{\eps^4} + \frac{2}{\eps^3}\ln\frac{-t}{\tau}
- \left(5\zeta_2 - \frac{1}{2}\ln^2\frac{-t}{\tau}\right)\frac{1}{\eps^2} 
- \left(\zeta_3+ 2\zeta_2\ln\frac{-t}{\tau}\right)\frac{1}{\eps} 
\nn\\ &-& {55\over 4}\zeta_4 +
\left( \zeta_2\zeta_3 - 41\zeta_5 + \zeta_4\ln\frac{-t}{\tau}
\right) \eps \nn\\ &-&\left( {95\over 2}\zeta_3^2 + {1695\over 8}\zeta_6
+ (18\zeta_2\zeta_3 + 42\zeta_5) \ln\frac{-t}{\tau} \right)
\eps^2 + \ord(\eps^3) \label{eq:2loopif}\\
&=& \frac{1}{2} \left[ C^{(1)}(t,\tau) \right]^2 + \frac{\zeta_2}{\eps^2}
+ \left(\zeta_3 + \zeta_2\ln\frac{-t}{\tau}\right)\frac{1}{\eps} \nn\\
&+& \left( \zeta_3\ln\frac{-t}{\tau} - 19\zeta_4\right)
+ \left( 4\zeta_4\ln\frac{-t}{\tau} - 2\zeta_2\zeta_3 - 39\zeta_5 \right) \eps \nn\\
&-&\left( 48 \zeta_3^2 + {1773\over 8}\zeta_6
+ (18\zeta_2\zeta_3 + 41\zeta_5) \ln\frac{-t}{\tau} \right)
\eps^2 + \ord(\eps^3)\, .\nn
\eea

Armed with this knowledge together with the two-loop amplitude, $m_5^{(2)}(\eps)$, one could extract the two-loop Lipatov vertex.  However, as we showed in Ref.~\cite{DelDuca:2008jg}, the Lipatov vertex satisfies its own iterative formula, and one can avoid needing to know $m_5^{(2)}(\eps)$.

\subsection{The two-loop five-point amplitude and the BDS ansatz}
\label{sec:iterative}

In the normalization of Refs.~\cite{DelDuca:2008pj,DelDuca:2008jg}, the iterative structure of
the two-loop five-point amplitude in the $\begin{cal}N\end{cal}=4$ super Yang-Mills theory
is given by~\cite{Anastasiou:2003kj,Bern:2005iz}
\beq
m_5^{(2)}(\eps)=\frac{1}{2}\,\left[m_5^{(1)}(\eps)\right]^2+\frac{2G^2(\eps)}{G(2\eps)}\,f^{(2)}(\eps)\,m_5^{(1)}(2\eps)+4Const^{(2)}+\ord(\eps)\, ,\label{eq:5ite}
\eeq
where $Const^{(2)}=-\zeta_2^2/2$, the $f^{(2)}$ function is
\beq
f^{(2)}(\eps)=-\zeta_2-\zeta_3\eps-\zeta_4\eps^2,
\eeq
and
\beq
G(\eps)=\frac{e^{-\gamma\eps}\Gamma(1-2\eps)}{\Gamma(1+\eps)\Gamma^2(1-\eps)}=1+\ord(\eps^2)\,
,\label{eq:geps}
\eeq
and where the one-loop five-point amplitude, $m_5^{(1)}(\eps)$, must be known through to $\ord(\eps^2)$.
In Ref.~\cite{Bern:2006vw}, the two-loop five-point amplitude has been shown to fulfil the BDS
ansatz by the numeric calculation of $m_5^{(1)}(\eps)$ through to $\ord(\eps^2)$ and of $m_5^{(2)}(\eps)$
though to finite terms.

In Ref.~\cite{DelDuca:2008pj}, the iterative structure~\cite{Anastasiou:2003kj,Bern:2005iz} and the Regge factorisation of the two-loop four-point amplitude have been used to
write the two-loop Regge trajectory and coefficient function through to finite terms in terms of the
constant $Const^{(2)}$, the function $f^{(2)}$, and of the one-loop coefficient function $C^{(1)}(\eps)$,
\beq\bsp\label{eq:ca2l}
\alpha^{(2)}(\eps)=&\, 2\, f^{(2)}(\eps)\, \alpha^{(1)}(2\eps) +\ord(\eps),\\
C^{(2)}(\eps)=&\,\frac{1}{2}\left[C^{(1)}(\eps)\right]^2
+\frac{2G^2(\eps)}{G(2\eps)}\,f^{(2)}(\eps)\,C^{(1)}(2\eps)+2\,Const^{(2)}+\ord(\eps),
\esp\eeq
where the one-loop coefficient function $C^{(1)}(\eps)$ is needed through to $\ord(\eps^2)$, and where we stress only the dependence on the dimensional-regularisation parameter $\eps$.

Combining Eq.~(\ref{eq:ca2l}), the iterative structure of the two-loop five-point amplitude~(\ref{eq:5ite}),
and the constant term of the high-energy expansion~(\ref{exp2loop}),
one can find an iteration formula for the two-loop gluon-production vertex~\cite{DelDuca:2008jg},
\beq
V^{(2)}(\eps)=\,\frac{1}{2}\left[V^{(1)}(\eps)\right]^2+\frac{2G^2(\eps)}{G(2\eps)}\,f^{(2)}(\eps)\,
V^{(1)}(2\eps)+\ord(\eps)\, ,\label{eq:bds}
\eeq
where the one-loop gluon-production vertex must be known through to $\ord(\eps^2)$. Note that in order to
compute the two-loop gluon-production vertex through to finite terms, it is not needed to know the two-loop
five-point amplitude or the two-loop coefficient function explicitly. It suffices to
know the one-loop five-point amplitude through to $\ord(\eps^2)$, from which one can derive
the one-loop gluon-production vertex to the same accuracy.

\subsection{Analytic continuation of the five-point amplitude to the physical region}
\label{sec:analytic}

We analytically continue the high-energy prescription for the 
colour-stripped amplitude (\ref{loop5pt}) to the physical region, where
$s, s_1, s_2$ are positive and $t_1, t_2$ are negative,
\beq
(-s) \to e^{-i\pi}\, s, \quad (-s_1) \to e^{-i\pi}\, s_1, \quad (-s_2) \to e^{-i\pi}\, s_2\, ,\label{eq:analsss}
\eeq
and where the multi-Regge kinematics are
\begin{equation}
s \gg s_{1},\ s_{2} \gg -t_1,\ -t_2\, .\label{eq:posmrk5pt} 
\end{equation}
\eqn{eq:analsss} implies that \eqns{exp1loop}{exp2loop} are continued
by $\ln(-s_j) = \ln(s_j) - i\pi$, for $s_j > 0$ and $j=1, 2$.
The mass-shell condition (\ref{mass}) and the analytic continuation (\ref{eq:analsss}) imply 
that the transverse scale $\kappa$ is continued as,
\beq
(-\kappa) \to e^{-i\pi}\, \kappa\, ,\label{eq:analk}
\eeq
and the mass-shell condition is reduced to the usual one in the physical region, \eqn{massnpt}.
Note that the expansions of Eqs.~(\ref{alphb})--(\ref{eq:coeffrescal}) are still valid, but because of
the analytic continuation on $\kappa$, which implies that
$\ln(-\kappa) = \ln(\kappa) - i\pi$, for $\kappa > 0$,
the gluon-production vertex becomes complex,
\beq
\bV^{(n)}(t_1,t_2,\kappa,\tau) = 
\left(\frac{\mu^2}{\kappa}\right)^{n\eps} 
V^{(n)}_{\rm phys}(t_1,t_2,\kappa,\tau)\, ,\label{eq:posrescal}
\eeq
with
\beq
V^{(n)}_{\rm phys}(t_1,t_2,\kappa,\tau) = 
e^{i\pi n\eps}\, V^{(n)}(t_1,t_2,e^{-i\pi}(-\kappa),\tau)\, .
\label{eq:vnlip}
\eeq

\section{The one-loop five-point amplitude}
\label{sec:1loop5ptamp}

We may write the one-loop five-point amplitude (\ref{eq:pent}) for general kinematics as,
\beq
m_5^{(1)}(1,2,3,4,5) = m_{5e}^{(1)}(1,2,3,4,5) + m_{5o}^{(1)}(1,2,3,4,5)\, ,\label{eq:5pt1leo}
\eeq
where the parity-even and odd contributions are given to all orders in $\eps$ 
by~\cite{Bern:1996ja,Bern:2006vw},
\bea
m_{5e}^{(1)}(1,2,3,4,5) &=& - \frac{1}{4}\, 2\, G(\eps)\,
\sum_{\rm cyclic} s_{12} s_{23} I_4^{1m}(1,2,3,45,\eps)\, 
,\label{eq:1l5pteven}\\
m_{5o}^{(1)}(1,2,3,4,5) &=& - \frac{\eps}{2}\, 2\, G(\eps)\,
\eps_{1234} I_5^{6-2\eps}(\eps)\, ,\label{eq:1l5ptodd}
\eea
where the cyclicity is over $i=1,\ldots,5$. Here $I_4^{1m}(1,2,3,45,\eps)$ is the one-mass box 
with the massive leg of virtuality $s_{45}$, $I_5^{6-2\eps}(\eps)$
is the pentagon evaluated in $6-2\eps$ dimensions, the contracted Levi-Civita tensor is
$\eps_{1234}= {\rm tr}[\gamma_5\!\!\not\!k_1\!\!\not\!\!k_2\!\not\!\!k_3\!\!\not\!k_4]$,
and we use the normalization of Refs.~\cite{DelDuca:2008pj,DelDuca:2008jg}, with $G(\eps)$
as in \eqn{eq:geps}.

For multi-Regge kinematics~(\ref{eq:mrk2}) in the Euclidean region~(\ref{eq:neginv}), 
the parity-even contribution is, to all orders in $\eps$,
\beq\bsp\label{eq:5pt2l}
m_{5e}^{(1)}&(1,2,3,4,5)\\
=&-\frac{1}{\eps^2}\,\muterm{-\kappa}\, \Gamma(1+\eps) \Gamma(1-\eps)\\
&+\frac{2}{\eps}\,\muterm{-t_1}\,\left(\psi(1)-\psi(-\eps)\right)+\frac{1}{\eps}\,\muterm{-t_2}\,\left(2\psi(1)-3\psi(-\eps)+\psi(1+\eps)\right)\\
&+\frac{1}{\eps^2}\,\muterm{-t_1}\,_2F_1\left(-\eps,1,1-\eps;\frac{t_1}{t_2}\right)-\frac{1}{\eps(1+\eps)}\,\muterm{-t_2}\,\frac{t_1}{t_2}\,_2F_1\left(1,1+\eps,2+\eps;\frac{t_1}{t_2}\right)\\
&+\frac{1}{\eps}\,\muterm{-t_1}\,\ln\frac{s_1}{\kappa}+\frac{1}{\eps}\,\muterm{-t_2}\,\ln\frac{s_2}{\kappa}\\
&+\frac{1}{\eps}\,\muterm{-t_1}\,\ln\frac{s_1}{t_2-t_1}+\frac{1}{\eps}\,\muterm{-t_2}\,\ln\frac{s_2}{t_2-t_1}\, ,
\esp\eeq
for $(-t_2)>(-t_1)$, and 
\beq\bsp
_2F_1\left(-\eps,1,1-\eps;z\right)=&\,1-\sum_{n=1}^\infty\,\li_n(z)\,\eps^n,\\
_2F_1\left(1,1+\eps,2+\eps;z\right)=&\,\frac{\li_1(z)}{z}+\left(\frac{\li_1(z)}{z}-\frac{\li_2(z)}{z}\right)\eps+\left(\frac{\li_3(z)}{z}-\frac{\li_2(z)}{z}\right)\eps^2\\
&+\left(\frac{\li_3(z)}{z}-\frac{\li_4(z)}{z}\right)\eps^3+\left(\frac{\li_5(z)}{z}-\frac{\li_4(z)}{z}\right)\eps^4+\ldots
\esp\eeq
where $\li_1(z) = -\ln(1- z)$. The parity-even term for $(-t_1)>(-t_2)$ is obtained
by exchanging $t_1$ and $t_2$ in \eqn{eq:5pt2l}.
\eqn{eq:5pt2l} is manifestly
real; it is also symmetric in $t_1$ and $t_2$, although not manifestly. It can be put in a
manifestly symmetric form, at the price of introducing imaginary parts, which cancel only
after combining all the terms. \eqn{eq:5pt2l} agrees through to $\ord(\eps^0)$ with the ${\cal N} = 4$
part of the one-loop five-gluon amplitude in QCD~\cite{Bern:1993mq} in the multi-Regge kinematics~\cite{DelDuca:1998cx}.

The parity-odd contribution is characterised by the contracted Levi-Civita tensor which can be written as,
\beq
\epslc=s_{12}s_{34}- s_{13}s_{24} + s_{14}s_{23} - 2\br{12}\sq{23}\br{34}\sq{41}\, .\label{eq:lct}
\eeq
In the multi-Regge kinematics~(\ref{eq:mrk2}) this becomes
\beq
\label{eq:LC}
\epslc=(-s)\left(p_{3\bot}p_{4\bot}^*-p_{4\bot}p_{3\bot}^*\right).
\eeq
Therefore we see that in the high energy limit, 
the parity-odd contribution (\ref{eq:1l5ptodd}) is given by
\beq
m_{5o}^{(1)}(1,2,3,4,5) = -\eps\, G(\eps)\, (-s)\,\left(p_{3\bot}p_{4\bot}^*-p_{4\bot}p_{3\bot}^*\right)\, 
\left(\frac{\mu^2}{-\kappa}\right)^\eps {\cal P}\, ,\label{eq:5pt1lo}
\eeq
where the function ${\cal P}$ is
\beq
{\cal P} = \left\{ 
\begin{array}{lll} \displaystyle\frac{1}{st_2}\, {\cal I}^{(IIa)}(\kappa,t_1,t_2) & \quad
\mbox{for \quad $-\sqrt{\frac{st_1}{s_1s_2}}+\sqrt{\frac{st_2}{s_1s_2}} > 1$} & \mbox{and $-t_1 < -t_2$}\, , \\ \\
\displaystyle\frac{1}{s_1s_2}\, {\cal I}^{(I)}(\kappa,t_1,t_2) & \quad
\mbox{for \quad $\sqrt{\frac{st_1}{s_1s_2}}+\sqrt{\frac{st_2}{s_1s_2}} < 1$}\, . & \end{array} \right.
\label{eq:forp}
\eeq
To all orders in $\eps$, ${\cal I}^{(IIa)}(\kappa,t_1,t_2)$ is~\cite{noi},
\beq\bsp\label{eq:PentNDRegIIa}
&{\cal I}^{(IIa)}(\kappa,t_1,t_2)\\
&=\, -{1\over\eps^3}\,y_2^{-\eps}\, \Gamma (1-2\epsilon )\, \Gamma (1+\epsilon )^2\, F_4\Big(1-2\eps, 1-\eps, 1-\eps, 1-\eps;-y_1, y_2\Big)\\
 &+\,
 {1\over \eps^3}\,
 \Gamma (1+\epsilon ) \,\Gamma (1-\epsilon ) \, F_4\Big(1, 1-\eps, 1-\eps, 1+\eps; -y_1, y_2\Big)\\
  &-\,{1\over \eps^2}\,y_1^\eps\,y_2^{-\eps}\,\Bigg\{\big[\ln y_1 + \psi(1-\eps) - \psi(-\eps)\big]\,F_4\big(1,1-\eps,1+\eps,1-\eps;-y_1, y_2\big)\\
&\qquad + {\partial\over \partial \delta}\,F^{2,1}_{0,2}\left(\begin{array}{cc|cccc|} 1+\delta & 1+\delta-\eps& 1&- & -& -\\
-&-&1+\delta&1-\eps& 1+\eps+\delta&-\end{array}\, -y_1, y_2\right)_{|\delta=0}\Bigg\}\\
&+\,{1\over \eps^2}\,y_1^{\eps}\,\Bigg\{\big[\ln y_1 + \psi(1+\eps) - \psi(-\eps)\big]\,F_4\big(1,1+\eps,1+\eps,1+\eps;-y_1, y_2\big)\\
&\qquad+{\partial\over \partial \delta}\,F^{2,1}_{0,2}\left(\begin{array}{cc|cccc|} 1+\delta & 1+\delta+\eps& 1&- & -& -\\
-&-&1+\delta&1+\eps& 1+\eps+\delta&-\end{array}\, -y_1, y_2\right)_{|\delta=0}\Bigg\},
\esp\eeq
with
\beq\label{eq:y12def}
y_1={\kappa\over t_2} {\rm ~~and~~} y_2={t_1\over t_2},
\eeq
and where we introduced the Appell function
\beq
F_4(a,b,c,d;x,y)\,=\sum_{m=0}^\infty\,\sum_{n=0}^\infty\,{\ph{a}{m+n}\,\ph{b}{m+n}\over\ph{c}{m}\ph{d}{n}}\,
{x^m\over m!}\,{y^n\over n!}\, ,\label{eq:appell}
\eeq
and the Kamp\'e de F\'eriet function
\beq
F^{p,q}_{p', q'}\left(\begin{array}{c|cc|}
\alpha_i & \beta_j &\gamma_j\\
\alpha'_k & \beta'_\ell &\gamma'_\ell
\end{array}\, x, y\right) = \sum_{m=0}^\infty \,\sum_{n=0}^\infty\,{\prod_i\,\ph{\alpha_i}{m+n}\,\prod_j\,\ph{\beta_j}{m}\,\ph{\gamma_j}{n}
\over \prod_k\,\ph{\alpha'_k}{m+n}\,\prod_\ell\,\ph{\beta'_\ell}{m}\,\ph{\gamma'_\ell}{n}}\,{x^m\over m!}\,{y^n\over n!},
\eeq
with $1\le i\le p$, $1\le j\le q$, $1\le k\le p'$ and $1\le \ell\le q'$. In Ref.~\cite{noi}, 
${\cal I}^{(IIa)}(\kappa,t_1,t_2)$ is given as a Laurent expansion through to $\ord(\eps)$ in terms of
Goncharov's multiple polylogarithms.

A few comments are in order: \eqn{eq:5pt1lo} starts at $\ord(\eps)$, because \eqn{eq:PentNDRegIIa} 
is finite: all the poles in $\eps$ cancel. Furthermore, because of the contracted Levi-Civita tensor (\ref{eq:lct}) 
in \eqn{eq:pent}, 
new momentum structures, other than the ones of \eqns{centrc}{lipeq} , occur in $m_5^{(1)}$.

In the region where $\sqrt{\frac{st_1}{s_1s_2}}-\sqrt{\frac{st_2}{s_1s_2}} > 1$ and $-t_1 > -t_2$,
which we term $IIb$, ${\cal I}^{(IIb)}$ is given by~\cite{noi},
\beq
{\cal I}^{(IIb)}(\kappa,t_1,t_2) = \frac{t_2}{t_1}\, {\cal I}^{(IIa)}(\kappa,t_2,t_1)
\eeq

In \eqn{eq:forp}, ${\cal I}^{(I)}(\kappa,t_1,t_2)$ can be derived from ${\cal I}^{(IIa)}(\kappa,t_1,t_2)$
by means of an analytic continuation, as detailed in Ref.~\cite{noi},
where it is also given explicitly to all orders in $\eps$,
as well as a Laurent expansion through to $\ord(\eps)$ in terms of
Goncharov's multiple polylogarithms.

\subsection{Soft limit}
\label{sec:sgl}

As discussed in Ref.~\cite{noi}, the 
limit in which the intermediate gluon becomes soft, $p_4 \rightarrow 0$, and thus $\kappa\rightarrow0$,
$t_1\to t$ and $t_2\to t$, is realised in the regions $IIa$ and $IIb$ of \eqn{eq:forp}. Thus,
\beq
\lim_{p_4\rightarrow0}
m_{5o}^{(1)}(1,2,3,4,5) = \eps\, G(\eps)\, \frac{p_{3\bot}p_{4\bot}^*-p_{4\bot}p_{3\bot}^*}{t}\,
{\cal I}^{(II)}(\kappa,t)\, .\label{eq:osoft}
\eeq
${\cal I}^{(II)}(\kappa,t)$ is obtained from \eqn{eq:PentNDRegIIa} by taking the limits
$t_1\to t$ and $t_2\to t$. Because ${\cal I}^{(II)}(\kappa,t)$ is logarithmic in $\kappa/t$, the parity-odd 
contribution is power suppressed and thus vanishes as $p_4\rightarrow 0$. 
Therefore the soft limit of the full one-loop five-point amplitude is given solely by the soft limit of 
the parity-even contribution,
\beq\bsp
\lim_{p_4\rightarrow0}&\,m_5^{(1)}(1,2,3,4,5)\\
=&\frac{2}{\eps}\,\muterm{-t}\,\left(\psi(1+\eps)-2\psi(-\eps)+\psi(1)+\ln\frac{s}{t}\right) -\frac{1}{\eps^2}\,\muterm{-\kappa}\,\frac{\pi\eps}{\sin(\pi\eps)}\, .\label{1l5ptsoft}
\esp\eeq
Using \eqn{eq:softm4}, we see that \eqn{1l5ptsoft} fulfills the soft limit of the one-loop five-point 
amplitude in the multi-Regge kinematics~(\ref{eq:softm5}).

\section{The one-loop gluon-production vertex}
\label{sec:onelooplip}

In order to compute the one-loop gluon-production vertex, 
we use \eqn{fullv} and subtract the one-loop trajectory (\ref{eq:alph1})
and coefficient function (\ref{eq:C1l})
from the one-loop five-point amplitude~(\ref{exp1loop}) and (\ref{eq:5pt2l}).
Thus, we obtain the gluon-production vertex,
\beq
{\bar V}^{(1)}(t_1,t_2,\tau,\kappa) = {\bar V}_e^{(1)}(t_1,t_2,\tau,\kappa) + 
{\bar V}_o^{(1)}(t_1,t_2,\kappa)\, ,\label{eq:V1leo} 
\eeq
in terms of parity-even and odd contributions,
\bea
{\bar V}_e^{(1)}(t_1,t_2,\tau,\kappa) &=& m_{5e}^{(1)}(1,2,3,4,5) - \balpha^{(1)}(t_1) L_1 - 
\balpha^{(1)}(t_2) L_2 - \bC^{(1)}(t_1,\tau) - \bC^{(1)}(t_2,\tau)\, ,\nn\\
{\bar V}_o^{(1)}(t_1,t_2,\kappa) &=& m_{5o}^{(1)}(1,2,3,4,5)\, .\label{eq:veo}
\eea
Because the high-energy coefficient functions and the Regge trajectory are parity-even,
the parity-odd part of the one-loop gluon-production vertex equals
the parity-odd part of the five-point amplitude (\ref{eq:5pt1lo}), and accordingly does not depend
on the factorisation scale $\tau$. Using \eqn{eq:coeffrescal},
in the unphysical region~(\ref{eq:neginv}) the parity-even contribution is, to all orders in $\eps$,
\beq\bsp\label{eq:V1l}
V_e^{(1)}&(t_1,t_2,\tau,\kappa)\\
=&-\frac{1}{\eps^2}\, \Gamma(1+\eps) \Gamma(1-\eps)\\
&+\left(\frac{\kappa}{t_1}\right)^\eps\left(\frac{\psi(1)-\psi(1+\eps)}{\eps}+
\frac{1}{\eps}\ln\frac{-t_1}{\tau}\right)
+\left(\frac{\kappa}{t_2}\right)^\eps\left(\frac{\psi(1)-\psi(-\eps)}{\eps}+
\frac{1}{\eps}\ln\frac{-t_2}{\tau}\right)\\
&+\frac{1}{\eps^2}\,\left(\frac{\kappa}{t_1}\right)^\eps\,_2F_1\left(\eps,1,1-\eps;\frac{t_1}{t_2}\right)-\frac{1}{\eps(1+\eps)}\,\left(\frac{\kappa}{t_2}\right)^\eps\,\frac{t_1}{t_2}\,_2F_1\left(1,1+\eps,2+\eps;\frac{t_1}{t_2}\right)\\
&-\frac{1}{\eps}\left[\left(\frac{\kappa}{t_1}\right)^\eps+\left(\frac{\kappa}{t_2}\right)^\eps\right]\,\left[\ln\frac{-\kappa}{\tau}+\ln\frac{t_1-t_2}{\tau}\right]\, .\esp\eeq
Using \eqns{eq:coeffrescal}{eq:5pt1lo}, the parity-odd contribution is, to all orders in $\eps$,
\beq
V_o^{(1)}(t_1,t_2,\kappa) = -\eps\, G(\eps)\, (-s)\,\left(p_{3\bot}p_{4\bot}^*-p_{4\bot}p_{3\bot}^*\right)\,
{\cal P}\, ,\label{eq:V1o}
\eeq
with ${\cal P}$ given by \eqn{eq:forp}, which also cancels the apparent dependence on $s$ above.

In the soft limit, $\kappa \rightarrow 0$, the parity-even part of the one-loop 
gluon-production vertex (\ref{eq:V1l}) agrees to all orders in $\eps$ with the soft limit of the corresponding QCD vertex~(\ref{eq:softv}). As we have seen in Section~\ref{sec:sgl}, the parity-odd part vanishes.

By expanding \eqn{eq:V1l} through to $\ord(\eps^2)$, and labelling the coefficients of the terms of $\ord(\eps)$ as
\bea
\vc_1(t_1,t_2) &=& \zeta_3 - \li_3\left(\frac{t_1}{t_2}\right) \nn\\
\vc_2(t_1,t_2,\tau) &=& \ln\frac{t_1}{t_2} \left( \li_2\left(\frac{t_1}{t_2}\right) + \zeta_2 \right) \nn\\
&& + \frac{1}{3} \ln^3\frac{-t_1}{\tau} - \frac{1}{2} \ln^2\frac{-t_1}{\tau} \ln\frac{-t_2}{\tau}
+ \frac{1}{6} \ln^3\frac{-t_2}{\tau} \nn\\
\vc_3(t_1,t_2,\tau) &=& \frac{1}{6} \ln^3\frac{-t_1}{\tau} \ln\frac{-t_2}{\tau} -
\frac{1}{8} \ln^4\frac{-t_1}{\tau} - \frac{1}{24} \ln^4\frac{-t_2}{\tau} \nn\\
&& - \frac{1}{2} \left( \ln^2\frac{-t_1}{\tau} - \ln^2\frac{-t_2}{\tau} \right)
\left( \li_2\left(\frac{t_1}{t_2}\right) + \zeta_2 \right)\, ,\label{eq:vcoeff} 
\eea
the even part of the one-loop Lipatov vertex becomes
\bea
\lefteqn{ V_e^{(1)}(t_1,t_2,\tau,\kappa) = -\frac{1}{\eps^2}\, \Gamma(1+\eps) \Gamma(1-\eps) } \nn\\
&& + \left[ \left(\frac{\kappa}{t_1}\right)^\eps+\left(\frac{\kappa}{t_2}\right)^\eps \right]
\left( -\frac{1}{\eps} \ln\frac{-\kappa}{\tau} + \eps\, \vc_1(t_1,t_2) \right) \nn\\
&& + \left( -\frac{1}{2} \ln^2\frac{t_1}{t_2} + \eps\, \vc_2(t_1,t_2,\tau) \right)
\left( \frac{-\kappa}{\tau} \right)^\eps + \eps^2\, \vc_3(t_1,t_2,\tau) + \ord(\eps^3)\, .\label{eq:eps2}
\eea
with the expansion of the first term given in \eqn{eq:sine}.

The expansion of \eqn{eq:V1o} through to $\ord(\eps^2)$ is provided by the expansion of the functions
${\cal I}^{(IIa)}(\kappa,t_1,t_2)$ and ${\cal I}^{(I)}(\kappa,t_1,t_2)$
of \eqn{eq:forp} through to $\ord(\eps)$
in terms of Goncharov's multiple polylogarithms~\cite{noi}.

\subsection{Analytic continuation of the one-loop vertex to the physical region}
\label{sec:analvert}

Using \eqn{eq:vnlip} and the prescription $\ln(-\kappa) = \ln(\kappa) - i\pi$, for $\kappa > 0$,
in the physical region where $s, s_1, s_2$ are positive and $t_1, t_2$ are negative,
the parity-even part of the one-loop gluon-production vertex (\ref{eq:V1l}) is
\beq\bsp\label{eq:V1lphys}
V^{(1)}_{e, \textrm{phys}}&(t_1,t_2,\tau,\kappa)\\
=&-\frac{1}{\eps^2}\, e^{i\pi\eps}\, \Gamma(1+\eps) \Gamma(1-\eps)\\
&+\left(\frac{\kappa}{-t_1}\right)^\eps\left(\frac{\psi(1)-\psi(1+\eps)}{\eps}+\frac{1}{\eps}\ln\frac{-t_1}{\tau}\right)
+\left(\frac{\kappa}{-t_2}\right)^\eps\left(\frac{\psi(1)-\psi(-\eps)}{\eps}+\frac{1}{\eps}\ln\frac{-t_2}{\tau}\right)\\
&+\frac{1}{\eps^2}\,\left(\frac{\kappa}{-t_1}\right)^\eps\,_2F_1\left(\eps,1,1-\eps;\frac{t_1}{t_2}\right)-\frac{1}{\eps(1+\eps)}\,\left(\frac{\kappa}{-t_2}\right)^\eps\,\frac{t_1}{t_2}\,_2F_1\left(1,1+\eps,2+\eps;\frac{t_1}{t_2}\right)\\
&-\frac{1}{\eps}\left[\left(\frac{\kappa}{-t_1}\right)^\eps+\left(\frac{\kappa}{-t_2}\right)^\eps\right]\,\left[\ln\frac{\kappa}{\tau}-i\pi+\ln\frac{t_1-t_2}{\tau}\right]\, .
\esp\eeq
Using the identity (\ref{eq:sine}) and
\beq
\pi\eps\,\frac{\cos(\pi\eps)}{\sin(\pi\eps)}=1+\eps(\psi(1-\eps)-\psi(1+\eps))\, ,\label{eq:gcos}
\eeq
the real part of the parity-even one-loop gluon-production vertex becomes
\beq\bsp\label{eq:V1lre}
\re V^{(1)}_{e, \textrm{phys}}&(t_1,t_2,\tau,\kappa)\\
=&-\frac{1+\eps(\psi(1-\eps)-\psi(1+\eps))}{\eps^2}\\
&+\left(\frac{\kappa}{-t_1}\right)^\eps\left(\frac{\psi(1)-\psi(1+\eps)}{\eps}+\frac{1}{\eps}\ln\frac{-t_1}{\tau}\right)
+\left(\frac{\kappa}{-t_2}\right)^\eps\left(\frac{\psi(1)-\psi(-\eps)}{\eps}+\frac{1}{\eps}\ln\frac{-t_2}{\tau}\right)\\
&+\frac{1}{\eps^2}\,\left(\frac{\kappa}{-t_1}\right)^\eps\,_2F_1\left(\eps,1,1-\eps;\frac{t_1}{t_2}\right)-\frac{1}{\eps(1+\eps)}\,\left(\frac{\kappa}{-t_2}\right)^\eps\,\frac{t_1}{t_2}\,_2F_1\left(1,1+\eps,2+\eps;\frac{t_1}{t_2}\right)\\
&-\frac{1}{\eps}\left[\left(\frac{\kappa}{-t_1}\right)^\eps+\left(\frac{\kappa}{-t_2}\right)^\eps\right]\,\left[\ln\frac{\kappa}{\tau}+\ln\frac{t_1-t_2}{\tau}\right],
\esp\eeq
which can readily expanded in $\eps$ like in \eqns{eq:vcoeff}{eq:eps2}.

The imaginary part is given by
\beq\label{eq:V1lim}
\im V^{(1)}_{e, \textrm{phys}}(t_1,t_2,\tau,\kappa)=
\frac{\pi}{\eps}\,\Bigg\{-1+\left(\frac{\kappa}{-t_1}\right)^\eps+\left(\frac{\kappa}{-t_2}\right)^\eps \Bigg\}.
\eeq

Taking into account the sign flip of the spin structure (\ref{eq:LC}), the analytic continuation of the
parity-odd part of the one-loop gluon-production vertex (\ref{eq:V1o}) is
\beq
V^{(1)}_{o, \textrm{phys}}(t_1,t_2,\tau,\kappa) = 
-\eps\, G(\eps)\, s\,\left(p_{3\bot}p_{4\bot}^*-p_{4\bot}p_{3\bot}^*\right)\, 
{\cal P}_{\textrm{phys}}\, ,\label{eq:v1lophys}
\eeq
where the function ${\cal P}_{\textrm{phys}}$ is,
\beq
{\cal P}_{\textrm{phys}} = \left\{ 
\begin{array}{lll} \displaystyle\frac{1}{st_2}\, {\cal I}^{(IIa)}_{\textrm{phys}}(\kappa,t_1,t_2) & \quad
\mbox{for \quad $-\sqrt{\frac{-st_1}{s_1s_2}}+\sqrt{\frac{-st_2}{s_1s_2}} > 1$} & \mbox{and $-t_1 < -t_2$}\, , \\ \\
\displaystyle\frac{1}{s_1s_2}\, {\cal I}^{(I)}_{\textrm{phys}}(\kappa,t_1,t_2) & \quad
\mbox{for \quad $\sqrt{\frac{-st_1}{s_1s_2}}+\sqrt{\frac{-st_2}{s_1s_2}} < 1$}\, . & \end{array} \right.
\label{eq:forphys}
\eeq
The analytic continuation (\ref{eq:analk}) implies that the ratios $y_1$ and $y_2$, \eqn{eq:y12def}, are continued as,
\beq
(-y_1) \to e^{-i\pi}\, y_1\,, \quad y_2\to y_2\, ,\label{eq:analy}
\eeq
and the functions ${\cal I}^{(I,II)}_{\textrm{phys}}(\kappa,t_1,t_2)$ are
continued according to \eqn{eq:vnlip}. Then \eqn{eq:PentNDRegIIa} is continued to,
\beq\bsp\label{eq:PentNDRegIIaphys}
&{\cal I}^{(IIa)}_{\textrm{phys}}(\kappa,t_1,t_2)\\
&=\, -e^{i\pi\eps}\, {1\over\eps^3}\,y_2^{-\eps}\, \Gamma (1-2\epsilon )\, \Gamma (1+\epsilon )^2\, 
F_4\Big(1-2\eps, 1-\eps, 1-\eps, 1-\eps;-y_1, y_2\Big)\\
 &+\, e^{i\pi\eps}\,
 {1\over \eps^3}\,
 \Gamma (1+\epsilon ) \,\Gamma (1-\epsilon ) \, F_4\Big(1, 1-\eps, 1-\eps, 1+\eps; -y_1, y_2\Big)\\
  &-\,{1\over \eps^2}\,(-y_1)^\eps\,y_2^{-\eps}\,\Bigg\{\big[\ln (-y_1) - i\pi + \psi(1-\eps) - \psi(-\eps)\big]\,F_4\big(1,1-\eps,1+\eps,1-\eps;-y_1, y_2\big)\\
&\qquad + {\partial\over \partial \delta}\,F^{2,1}_{0,2}\left(\begin{array}{cc|cccc|} 1+\delta & 1+\delta-\eps& 1&- & -& -\\
-&-&1+\delta&1-\eps& 1+\eps+\delta&-\end{array}\, -y_1, y_2\right)_{|\delta=0}\Bigg\}\\
&+\,{1\over \eps^2}\,(-y_1)^{\eps}\,\Bigg\{\big[\ln (-y_1) - i\pi + \psi(1+\eps) - \psi(-\eps)\big]\,F_4\big(1,1+\eps,1+\eps,1+\eps;-y_1, y_2\big)\\
&\qquad+{\partial\over \partial \delta}\,F^{2,1}_{0,2}\left(\begin{array}{cc|cccc|} 1+\delta & 1+\delta+\eps& 1&- & -& -\\
-&-&1+\delta&1+\eps& 1+\eps+\delta&-\end{array}\, -y_1, y_2\right)_{|\delta=0}\Bigg\},
\esp\eeq
where the Appell and the Kamp\'e de F\'eriet functions stay real in the analytic continuation~\cite{noi}.
Like in the Euclidean region in \sec{sec:1loop5ptamp}, in \eqn{eq:PentNDRegIIaphys} all the poles in
$\eps$ cancel.
\eqn{eq:PentNDRegIIaphys} may be expanded in $\eps$ in terms of real ${\cal M}$ functions, introduced
in Ref.~\cite{noi}. It could also be expanded in terms of Goncharov's multiple polylogarithms, to the price,
though, of introducing a complicated and fictitious analytic structure: the Goncharov's polylogarithms
would occur with several spurious imaginary parts, which ultimately would have to cancel in order to respect
the fact that the Appell and the Kamp\'e de F\'eriet functions are real.

Using the identities (\ref{eq:gcos}) and (\ref{eq:sine}), the real part of the function 
$I^{(IIa)}_{\textrm{phys}}(\kappa,t_1,t_2)$ is
\beq\bsp\label{eq:PentNDRegIIaphysre}
&\re {\cal I}^{(IIa)}_{\textrm{phys}}(\kappa,t_1,t_2)\\
&=\, - y_2^{-\eps}\, \frac{1+\eps(\psi(1-\eps)-\psi(1+\eps))}{\eps^3}\,
\frac{\Gamma (1-2\epsilon )\, \Gamma (1+\epsilon )}{\Gamma (1-\epsilon )}
F_4\Big(1-2\eps, 1-\eps, 1-\eps, 1-\eps;-y_1, y_2\Big)\\
 &+\, \frac{1+\eps(\psi(1-\eps)-\psi(1+\eps))}{\eps^3}\,
 F_4\Big(1, 1-\eps, 1-\eps, 1+\eps; -y_1, y_2\Big)\\
  &-\,(-y_1)^\eps\,y_2^{-\eps}\, {1\over \eps^2}\, \Bigg\{\big[\ln (-y_1) + \psi(1-\eps) - \psi(-\eps)\big]\,F_4\big(1,1-\eps,1+\eps,1-\eps;-y_1, y_2\big)\\
&\qquad + {\partial\over \partial \delta}\,F^{2,1}_{0,2}\left(\begin{array}{cc|cccc|} 1+\delta & 1+\delta-\eps& 1&- & -& -\\
-&-&1+\delta&1-\eps& 1+\eps+\delta&-\end{array}\, -y_1, y_2\right)_{|\delta=0}\Bigg\}\\
&+\, (-y_1)^{\eps}\, {1\over \eps^2}\,
\Bigg\{\big[\ln (-y_1) + \psi(1+\eps) - \psi(-\eps)\big]\,F_4\big(1,1+\eps,1+\eps,1+\eps;-y_1, y_2\big)\\
&\qquad+{\partial\over \partial \delta}\,F^{2,1}_{0,2}\left(\begin{array}{cc|cccc|} 1+\delta & 1+\delta+\eps& 1&- & -& -\\
-&-&1+\delta&1+\eps& 1+\eps+\delta&-\end{array}\, -y_1, y_2\right)_{|\delta=0}\Bigg\},
\esp\eeq
which, if desired, may be expanded in $\eps$ in terms of real ${\cal M}$ functions.

The imaginary part is
\beq\bsp\label{eq:PentNDRegIIaphysim}
&\im {\cal I}^{(IIa)}_{\textrm{phys}}(\kappa,t_1,t_2) \\
&=\, {\pi\over\eps^2}\, \Bigg\{ - y_2^{-\eps}\, 
\frac{\Gamma (1-2\epsilon )\, \Gamma (1+\epsilon )}{\Gamma (1-\epsilon )}\,
F_4\Big(1-2\eps, 1-\eps, 1-\eps, 1-\eps;-y_1, y_2\Big)\\
 &\qquad\quad + F_4\Big(1, 1-\eps, 1-\eps, 1+\eps; -y_1, y_2\Big)\\
 &\qquad\quad + (-y_1)^\eps\,y_2^{-\eps}\, F_4\big(1,1-\eps,1+\eps,1-\eps;-y_1, y_2\big)\\
 &\qquad\quad - (-y_1)^{\eps}\, F_4\big(1,1+\eps,1+\eps,1+\eps;-y_1, y_2\big) \Bigg\}\, .
\esp\eeq
The Appell functions in Eq.~(\ref{eq:PentNDRegIIaphysim}) are all reducible to Gauss' hypergeometric function. We find
\beq\bsp
&\im {\cal I}^{(IIa)}_{\textrm{phys}}(\kappa,t_1,t_2) \\
&=\, {\pi\over\eps^2}\,{1\over\sqrt{\lambda(1,-y_1,y_2)}}\, \Bigg\{ - y_2^{-\eps}\, 
\frac{\Gamma (1-2\epsilon )\, \Gamma (1+\epsilon )}{\Gamma (1-\epsilon )}\,
\lambda(1,-y_1,y_2)^\eps\\
 &\qquad\quad + {_2F_1}\left(1,2\eps,1+\eps;{(1-\lambda_1)\,\lambda_2\over 1-\lambda_1\,\lambda_2}\right)\\
 &\qquad\quad + (-y_1)^\eps\,y_2^{-\eps}\, {_2F_1}\left(1,2\eps,1+\eps;{\lambda_1\,(1-\lambda_2)\over 1-\lambda_1\,\lambda_2}\right)\\
 &\qquad\quad - (-y_1)^{\eps}\, {_2F_1}\left(1,2\eps,1+\eps;{\lambda_1\,\lambda_2\over 1-\lambda_1\,\lambda_2}\right)\Bigg\}\, ,
\esp\eeq
where $\lambda$ denotes the K\"allen function, $\lambda(x,y,z) = x^2+y^2+z^2-2xy-2xz-2yz$, and
\bea
&\lambda_1 = \displaystyle{-1\over 2y_1}(y_2-y_1-1+\sqrt{\lambda(1,-y_1,y_2)})\, \nn\\
&\lambda_2 = \displaystyle{1\over 2y_2}(y_2-y_1-1+\sqrt{\lambda(1,-y_1,y_2)}).
\eea
The hypergeometric function can be expanded into a Taylor series in $\eps$,
\beq
{_2F_1}\left(1,2\eps,1+\eps;z\right) = 1-2\,\eps\,\ln(1-z)+ 
2\,\eps^2\,\left({1\over2}\ln^2(1-z) - \textrm{Li}_2(z)\right)+\ord(\eps^3).
\eeq

\section{The two-loop gluon-production vertex}
\label{sec:twolooplip}

In terms of parity-even and odd contributions, the two-loop gluon-production vertex is
\beq
V^{(2)}(t_1,t_2,\tau,\kappa) = V_e^{(2)}(t_1,t_2,\tau,\kappa) + V_o^{(2)}(t_1,t_2,\tau,\kappa)\, 
.\label{eq:V2leo} 
\eeq
Using \eqn{eq:V1leo} and the iteration formula (\ref{eq:bds}), \eqn{eq:V2leo} becomes
\bea
V_e^{(2)}(\eps)&=&\,\frac{1}{2}\left[V_e^{(1)}(\eps)\right]^2 + \frac{2G^2(\eps)}{G(2\eps)}\,f^{(2)}(\eps)\,
V_e^{(1)}(2\eps)+\ord(\eps)\, ,\label{eq:bdse}\\
V_o^{(2)}(\eps)&=& V_e^{(1)}(\eps) V_o^{(1)}(\eps) + \ord(\eps)\, ,\label{eq:bdso}
\eea
where the parity-even and odd parts of the one-loop gluon-production vertex must be known 
through to $\ord(\eps^2)$.
We used the fact that $V_o^{(1)}(\eps) = \ord(\eps)$, so it does not contribute to
the square of the one-loop vertex in \eqn{eq:bdse}, and to the term proportional to $f^{(2)}$
in \eqn{eq:bdso}.

In the unphysical Euclidean region~(\ref{eq:neginv}), \eqn{eq:bdse} becomes
\bea
\lefteqn{ V_e^{(2)}(t_1,t_2,\tau,\kappa) } \nn\\
&=& \frac{1}{2\eps^4} \left(\frac{\pi\eps}{\sin(\pi\eps)} \right)^2
- \frac{2G^2(\eps)}{G(2\eps)}\,f^{(2)}(\eps)\, \frac{1}{4\eps^2} \frac{2\pi\eps}{\sin(2\pi\eps)} \nn\\
&+& \frac{1}{8} \ln^4\frac{t_1}{t_2} - \vc_3(t_1,t_2,\tau) + \zeta_2 \ln^2\frac{t_1}{t_2} \nn\\
&+& \left[ \left(\frac{\kappa}{t_1}\right)^{2\eps}+\left(\frac{\kappa}{t_2}\right)^{2\eps} \right]
\left[ \frac{1}{2\eps^2} \ln^2\frac{-\kappa}{\tau} - \left( \frac{1}{\eps} f^{(2)}(\eps)
+ \vc_1(t_1,t_2) \right) \ln\frac{-\kappa}{\tau} \right] \nn\\
&+& \left(\frac{\kappa}{t_1}\right)^\eps \left(\frac{\kappa}{t_2}\right)^\eps
\left( \frac{1}{\eps^2} \ln^2\frac{-\kappa}{\tau} - 2\ln\frac{-\kappa}{\tau} \vc_1(t_1,t_2) \right) \nn\\
&-& \left[ \left(\frac{\kappa}{t_1}\right)^\eps+\left(\frac{\kappa}{t_2}\right)^\eps \right]
\frac{1}{\eps^2} \frac{\pi\eps}{\sin(\pi\eps)}
\left( -\frac{1}{\eps} \ln\frac{-\kappa}{\tau} + \eps\vc_1(t_1,t_2) \right) \nn\\
&-& \left[ \left(\frac{\kappa}{t_1}\right)^\eps+\left(\frac{\kappa}{t_2}\right)^\eps \right]
\left( \frac{-\kappa}{\tau} \right)^\eps \frac{1}{\eps} \ln\frac{-\kappa}{\tau}
\left( -\frac{1}{2} \ln^2\frac{t_1}{t_2} + \eps\, \vc_2(t_1,t_2,\tau) \right) \nn\\
&-& \left( \frac{-\kappa}{\tau} \right)^\eps \frac{1}{\eps^2} \frac{\pi\eps}{\sin(\pi\eps)}
\left( -\frac{1}{2} \ln^2\frac{t_1}{t_2} + \eps\, \vc_2(t_1,t_2,\tau) \right)
+ \ord(\eps) \, ,\label{eq:lipvert2}
\eea
where we have used \eqn{eq:eps2} and collected the terms according to the different analytic structures.

The parity-odd part of the two-loop gluon-production vertex, $V_o^{(2)}(t_1,t_2,\tau,\kappa)$,
starts at $\ord(\eps^{-1})$ and is given by the product of \eqns{eq:V1l}{eq:V1o}.

The analytic continuation of the two-loop gluon-production vertex to the physical region where 
$s, s_1, s_2$ are positive and $t_1, t_2$ are negative, may be performed as in \sec{sec:analvert}.

\section{Conclusions}
\label{sec:concl}

In this paper we have computed the one-loop five-point amplitude $m_5^{(1)}$ in the planar $\begin{cal}N\end{cal}=4$ 
supersymmetric Yang-Mills theory in the multi-Regge kinematics, using the
calculation of the one-loop pentagon in $D=6-2\eps$ performed in a companion paper~\cite{noi}.
We have presented  $m_5^{(1)}$ in the Euclidean region (\ref{eq:mrk2})
as an expression to all orders in $\eps$ in terms of parity-even, \eqn{eq:5pt2l}, and parity-odd contributions,
\eqn{eq:5pt1lo}, starting at $\ord(\eps^{-2})$
and at $\ord(\eps)$, respectively.

Using the high-energy factorisation for colour-stripped amplitudes, we have computed 
the one-loop gluon-production vertex to all orders in $\eps$ in \eqns{eq:V1l}{eq:V1o}.
Because the high-energy coefficient functions and the Regge trajectory are parity-even,
the parity-odd part of the one-loop gluon-production vertex equals the parity-odd part of the
one-loop five-point amplitude, and thus appears at $\ord(\eps)$.
The Laurent expansion in $\eps$ through to $\ord(\eps^2)$ is given in \eqn{eq:eps2} 
for the parity-even part, and in Ref.~\cite{noi} for the parity-odd part
in terms of Goncharov's multiple polylogarithms. In \eqns{eq:V1lphys}{eq:v1lophys},
we have continued analytically the all-orders-in-$\eps$
one-loop gluon-production vertex to the physical region.
The even-parity part may be easily expanded in $\eps$; the odd-parity part
may be more conveniently expanded in $\eps$ in terms of real ${\cal M}$ functions, 
as in Ref.~\cite{noi}.

The iterative structure of the two-loop five-point amplitude implied by the BDS 
ansatz, together with the high-energy factorisation, implies
an iterative structure of the gluon-production vertex. Thus, the knowledge of the
one-loop gluon-production vertex through to $\ord(\eps^2)$, allows us to perform the first computation
of the two-loop gluon-production vertex through to finite terms, which we present in \eqn{eq:lipvert2}
as an expansion starting at $\ord(\eps^{-4})$. The parity-odd part of the two-loop 
gluon-production vertex appears at $\ord(\eps^{-1})$ and is given by the product 
of \eqns{eq:V1l}{eq:V1o}\footnote{In Ref.~\cite{Bartels:2008ce}
the logarithm of the gluon-production vertex has been introduced. If exponentiated, it yields
at two-loop order
the poles in $\eps$ through to $\ord(\eps^{-2})$, but it misses the single poles in $\eps$, as well
as the finite terms. Accordingly, it lacks completely the parity-odd contribution.}.

If augmented by the soft-limit contribution to $\ord(\eps)$, which is as yet unknown,
the two-loop gluon-production vertex could be used as one of the building blocks of the kernel
of a BFKL equation at next-to-next-to-leading logarithmic (NNLL) accuracy. The other building blocks
are the three-loop Regge 
trajectory~\cite{DelDuca:2008pj,Bartels:2008ce,Drummond:2007aua,Naculich:2007ub},
the one-loop vertex for the emission of two gluons along the ladder (computed in~\cite{Bartels:2008ce}
only for two gluons of the same helicity) and the tree vertex for the emission of three gluons 
along the ladder~\cite{Del Duca:1999ha,Antonov:2004hh}.

\section*{Acknowledgements}

CD and VDD thank the IPPP Durham and EWNG and CD thank the LNF
Frascati for the warm hospitality at various stages of this work. CD is a
research fellow of the \emph{Fonds National de la Recherche Scientifique},
Belgium. This work was partly supported by MIUR under contract 2006020509$_0$04,
and by the EC Marie-Curie Research Training Network ``Tools and Precision
Calculations for Physics Discoveries at Colliders'' under contract
MRTN-CT-2006-035505. EWNG gratefully acknowledges the support of the Wolfson
Foundation and the Royal Society.

\appendix

\section{Multi-parton kinematics}
\label{sec:mpk}

We consider the production of three gluons of outgoing momentum $p_i$, with 
$i=3,...,5$ in the scattering between two gluons of ingoing 
momenta $p_1$ and $p_2$\footnote{By convention
we consider the scattering in the unphysical region where all momenta 
are taken as outgoing, and then we analitically continue to the
physical region where $p_1^0<0$ and $p_2^0<0$.}.

Using light-cone coordinates $p^{\pm}= p_0\pm p_z $, and
complex transverse coordinates $p_{\perp} = p^x + i p^y$, with scalar
product $2 p\cdot q = p^+q^- + p^-q^+ - p_{\perp} q^*_{\perp} - p^*_{\perp} 
q_{\perp}$, the 4-momenta are,
\begin{eqnarray}
p_2 &=& \left(p_2^+/2, 0, 0,p_2^+/2 \right) 
    \equiv  \left(p_2^+ , 0; 0, 0 \right)\, ,\nonumber \\
p_1 &=& \left(p_1^-/2, 0, 0,-p_1^-/2 \right) 
    \equiv  \left(0, p_1^-; 0, 0\right)\, ,\label{in}\\
p_i &=& \left( (p_i^+ + p_i^- )/2, 
               {\rm Re}[p_{i\perp}],
               {\rm Im}[p_{i\perp}], 
               (p_i^+ - p_i^- )/2 \right)\nonumber\\
   &\equiv& \left(|p_{i\perp}| e^{y_i}, |p_{i\perp}| e^{-y_i}; 
|p_{i\perp}|\cos{\phi_i}, |p_{i\perp}|\sin{\phi_i}\right)\, \,,\nonumber
\end{eqnarray}
where $y$ is the rapidity. The first notation above is the 
standard representation 
$p^\mu =(p^0,p^x,p^y,p^z)$, while in the second we have the + and -
components on the left of the semicolon, 
and on the right the transverse components.
In the following, if not  differently stated, $p_i$ and $p_j$ are always 
understood to lie in the range $3\le i,j \le n$. The mass-shell condition is
$|p_{i\perp}|^2 = p_i^+ p_i^-$. Using momentum conservation,
\beq
0 = \sum_{i=3}^5 p_{i\perp}\, ,\qquad
p_2^+ = -\sum_{i=3}^5 p_i^+\, ,\qquad
p_1^- = -\sum_{i=3}^5 p_i^-\, ,\label{nkin}
\eeq
the Mandelstam invariants may be written as,
\begin{eqnarray}
s_{ij} &=& 2 p_i\cdot p_j = p_i^+ p_j^- + p_i^- p_j^+
- p_{i\perp} p_{j\perp}^* - p_{i\perp}^* p_{j\perp}\, ,\label{eq:mandelst}
\end{eqnarray}
so that 
\begin{eqnarray}
s &=& 2 p_1\cdot p_2 = \sum_{i,j=3}^5 p_i^+ p_j^-\, , \nonumber\\ 
s_{2i} &=& 2 p_2\cdot p_i = -\sum_{j=3}^5 p_i^- p_j^+\, , \label{inv}\\ 
s_{1i} &=& 2 p_1\cdot p_i = -\sum_{j=3}^5 p_i^+ p_j^-\, . \nonumber
\end{eqnarray}

Using the spinor representation of Ref.~\cite{Del Duca:1999ha},
\begin{equation} \begin{array}{cc}
\psi_+(p_i) = \left( \begin{array}{c} \sqrt{p_i^+}\\ \sqrt{p_i^-} e^{i\phi_i}\\ 0\\ 0\end{array}\right)\, ,
& \psi_-(p_i) = \left(
\begin{array}{c} 0\\ 0\\ \sqrt{p_i^-} e^{-i\phi_i}\\ 
-\sqrt{p_i^+}\end{array}\right)\, ,
\\ \\ \psi_+(p_2) = i \left( \begin{array}{c} \sqrt{-p_2^+}\\ 0\\
0\\ 0\end{array} \right)\, , & \psi_-(p_2) = i \left( \begin{array}{c}
0\\ 0\\ 0\\ -\sqrt{-p_2^+} \end{array}\right)\, ,\\ \\ 
\psi_+(p_1) = -i
\left( \begin{array}{c} 0\\ \sqrt{-p_1^-} \\ 0\\ 0\end{array}\right)\, , &
\psi_-(p_1) = -i \, \left( \begin{array}{c} 0\\ 0\\
\sqrt{-p_1^-}\\ 0\end{array}\right)\, . \end{array}\label{spin}
\end{equation}
for the momenta (\ref{in})\footnote{The spinors of the 
incoming partons must be continued
to negative energy after the complex conjugation, \emph{e.g.}
$\overline{\psi_{+}(p_2)}= i \left( \sqrt{-p_2^+}, 0, 0, 0 \right)$.},
the spinor products are
\begin{eqnarray}
\langle 2 1\rangle 
&=& -\sqrt{s}\, ,\nonumber\\
\langle 2 i\rangle &=& - i \sqrt{\frac{-p_2^+}{p_i^+}}\, p_{i\perp}\, ,\label{spro}\\ 
\langle i 1\rangle &=& i \sqrt{-p_1^- p_i^+}\, ,\nonumber\\ 
\langle i j\rangle &=& p_{i\perp}\sqrt{\frac{p_j^+}{p_i^+}} - p_{j\perp}
\sqrt{\frac{p_i^+}{ p_j^+}}\, , \nonumber
\end{eqnarray}
where we have used the mass-shell condition $|p_{i\perp}|^2 = p_i^+ p_i^-$.
%
%

\section{Multi-Regge kinematics}
\label{sec:mrk}

In the multi-Regge kinematics, we require that the gluons
are strongly ordered in rapidity and have comparable transverse momentum
(\ref{mrknpt}).
This is equivalent to require a strong ordering of the light-cone coordinates,
\begin{equation}
p_3^+\gg p_4^+\gg p_5^+; \qquad p_3^-\ll p_4^-\ll p_5^-.
\end{equation}
Momentum conservation (\ref{nkin}) then becomes
\beq
0 = \sum_{i=3}^5 p_{i\perp}\, ,\qquad
p_2^+ \simeq -p_3^+\, ,\qquad
p_1^- \simeq -p_5^-\, ,\label{mrkin}
\eeq
where the $\simeq$ sign is understood to mean ``equals up to corrections
of next-to-leading accuracy''.
The Mandelstam invariants (\ref{inv}) are reduced to,
\begin{eqnarray}
s &=& 2 p_1\cdot p_2 \simeq p_3^+ p_5^-, \nonumber\\ 
s_{2i} &=& 2 p_2\cdot p_i \simeq - p_3^+ p_i^-, \label{mrinv}\\ 
s_{1i} &=& 2 p_1\cdot p_i \simeq - p_i^+ p_5^-, \nonumber\\ 
s_{ij} &=& 2 p_i\cdot p_j \simeq p_i^+ p_j^-\qquad i < j\, .\nonumber
\end{eqnarray}
The product of the two successive invariants $s_{34}$ and $s_{45}$ fixes the mass shell of gluon 4,
\beq
s_{34}s_{45} \simeq p_3^+ p_4^- p_4^+ p_5^- = 
|p_{4\perp}|^2 p_3^+ p_5^- \simeq |p_{4\perp}|^2 s\, .\nn
\eeq
Thus,
\beq
|p_{4\perp}|^2 = \frac{s_{34}s_{45}}{s}\, .\label{eq:masshell}
\eeq

The spinor products (\ref{spro}) are,
\begin{eqnarray}
\langle 2 1\rangle &\simeq& -\sqrt{p_3^+ p_5^-}\, ,\nonumber\\
\langle 2 i\rangle &\simeq& - i\sqrt{\frac{p_3^+}{ p_i^+}}\, p_{i\perp}\, ,\label{mrpro}\\
\langle i 1\rangle &\simeq& i\sqrt{p_i^+ p_5^-}\, ,\nonumber\\ 
\langle i j\rangle &\simeq& -\sqrt{\frac{p_i^+}{p_j^+}}\,
p_{j\perp}\, \qquad {\rm for}\, y_i>y_j \, .\nonumber
\end{eqnarray}

\section{The soft limit of the one-loop five-point amplitude}
\label{sec:appc}

We consider the five--point amplitude of Sect.~\ref{sec:5pthel}, and take the limit where
the intermediate gluon becomes soft, $p_4\to 0$. In this limit, the one-loop five-point
amplitude factorises as~\cite{Bern:1998sc,Kosower:1999rx},
\bea
\lim_{p_4\to 0} m_5^{\rm 1-loop}(1,2,3,4^\lambda,5) &=& 
{\rm Soft}^{\rm tree}(3,4^\lambda,5)\, m_4^{\rm 1-loop}(1,2,3,5)
\nn\\ &+& {\rm Soft}^{\rm 1-loop}(3,4^\lambda,5)\, m_4^{\rm tree}(1,2,3,5)
\eea
where the one-loop soft-gluon function, to all orders of $\eps$, is
\beq
{\rm Soft}^{\rm 1-loop}(3,4^\lambda,5) = - \bar{g}^2\, \frac{1}{\eps^2}\,\frac{\pi\eps}{\sin(\pi\eps)}
\left( \frac{\mu^2 (-s_{35})}{(-s_{34}) (-s_{45})}\right)^\eps {\rm Soft}^{\rm tree}(3,4^\lambda,5)
\eeq
and the tree-level soft function is
\beq
{\rm Soft}^{\rm tree}(3,4^+,5) = \frac{\langle 3 5\rangle}{\langle 3 4\rangle \langle 4 5\rangle}\, .
\eeq
For the MHV amplitude we are considering, the soft limit for a negative helicity gluon is trivial
and is obtained from this by helicity reversal. In the multi-Regge kinematics (\ref{eq:mrk2}),
$s_{35}=s$; then, using the on-shell condition (\ref{mass}) and the normalisation of \eqn{elasexpand},
the soft-gluon limit of the one-loop five-point coefficient becomes
\beq
\lim_{p_4\to 0} m_5^{(1)}(1,2,3,4,5) = m_4^{(1)}(1,2,3,5) - \frac{1}{\eps^2}\,\frac{\pi\eps}{\sin(\pi\eps)}
\left(\frac{\mu^2}{-\kappa}\right)^\eps\, ,\label{eq:softm5}
\eeq
where we have factored out
\beq
\lim_{p_4\to 0} m_5^{(0)}(1,2,3,4^\lambda,5) = m_4^{(0)}(1,2,3,5) {\rm Soft}^{\rm tree}(3,4^\lambda,5)\, .
\eeq
In addition, in the soft limit of gluon 4, we can write the one-loop coefficient (\ref{exp1loop}) as,
\beq
\lim_{\kappa\to 0} m_5^{(1)}(1,2,3,4,5) = m_4^{(1)}(1,2,3,5) 
+ \bar{\alpha}^{(1)}(t) \ln\frac{-\kappa}{\tau} 
+ \lim_{\kappa\to 0} \bV^{(1)}(t,t,\tau,\kappa)\, ,\label{eq:softka}
\eeq
with $t_1=t_2=t$ and
\beq
m_4^{(1)} = \bar{\alpha}^{(1)}(t) \ln\frac{-s}{\tau} + 2\, \bC^{(1)}(t,\tau)\, .\label{eq:softm4}
\eeq
Equating \eqns{eq:softm5}{eq:softka} and using \eqns{eq:coeffrescal}{eq:alph1}, we obtain
the soft limit of the one-loop gluon-production vertex, to all orders of $\eps$~\cite{DelDuca:1998cx},
\beq
\lim_{\kappa\to 0} V^{(1)}(t,t,\tau,\kappa) = - \frac{1}{\eps^2}\,\frac{\pi\eps}{\sin(\pi\eps)}
- \frac{2}{\eps} \left( \frac{\kappa}{t}\right)^\eps \ln\frac{-\kappa}{\tau}\, ,\label{eq:softv}
\eeq
with
\bea
\frac{\pi\eps}{\sin(\pi\eps)} &=& \Gamma(1+\eps) \Gamma(1-\eps) = 1 + \zeta_2\eps^2
+ \frac{7}{4} \zeta_4\eps^4 + \frac{31}{16} \zeta_6\eps^6 + \cdots \nn\\
&=& \sum_{n=0}^{\infty} c_n\eps^{2n}, \quad {\rm with} \quad c_0=1, \quad 
c_n=\frac{2^{2n-1}-1}{2^{2(n-1)}} \zeta_{2n}\, .\label{eq:sine}
\eea

\end{document}